\documentclass[acmtog, authorversion, nonacm]{acmart}
\usepackage{multirow}
\usepackage{colortbl}
\usepackage{enumitem}
\usepackage{subcaption}
\usepackage{adjustbox}
\usepackage{caption}
\usepackage{booktabs}
\usepackage{balance}

\AtBeginDocument{%
  \providecommand\BibTeX{{%
    \normalfont B\kern-0.5em{\scshape i\kern-0.25em b}\kern-0.8em\TeX}}}

\begin{document}

\title{SaludConectaMX: Lessons Learned from Deploying a Cooperative Mobile Health System for Pediatric Cancer Care in Mexico} %Health Monitoring System?
%{HealthApp-MD System: Bridging Social-Clinical Gaps to Improve the Care of Children with Cancer in Mexico}
%{HealthApp-MD System: A Collaborative Socioclinical Telehealth Tool to Address the Evolving Healthcare Needs of Children with Cancer in Mexico}
\renewcommand{\shorttitle}{SaludConectaMX: Lessons Learned}%{HealthApp-MD System: A Telehealth Solution}

\author{Jennifer J. Schnur}
\orcid{0000-0001-5710-6795}
%\email{jschnur@nd.edu}
\authornotemark[1]

\author{Angélica Garcia-Martínez}
\orcid{0000-0003-2735-7395}
%\email{agarci35@nd.edu}
\authornotemark[1]

\author{Patrick Soga}
%\email{psoga117@gmail.com}%{psoga@nd.edu}
\orcid{0000-0001-8585-9241}
\authornote{Authors contributed equally to this paper.}

\author{Karla Badillo-Urquiola}
\orcid{0000-0002-1165-3619}
%\email{kbadillou@nd.edu}
\authornotemark[1]

\author{Alejandra J. Botello}
\orcid{0009-0003-8235-5752}
%\email{abotello@nd.edu}

\author{Ana Calderon Raisbeck}
\orcid{0009-0003-7216-9796}
%\email{anacalderonraisbeck@gmail.com}%{acalder4@nd.edu}

\author{Sugana Chawla}
\orcid{0009-0001-7530-5189}
%\email{schawla@nd.edu}

\author{Josef Ernst}
\orcid{0009-0003-3416-9154}
%\email{jernst981@gmail.com}%{jernst2@nd.edu}

\author{William Gentry}
\orcid{0009-0000-3012-5902}
%\email{william.barr.gentry@gmail.com}%{wgentry@nd.edu}
  
\author{Richard P. Johnson}
\orcid{0000-0002-1550-6325}
%\email{rick.johnson@nd.edu}

\author{Michael Kennel}
\orcid{0000-0002-6939-3207}
%\email{mkennel@nd.edu}

\author{Jesús Robles}
\orcid{0009-0007-2827-415X}
%\email{jrobles0302@gmail.com}%{jrobles2@nd.edu}

\author{Madison Wagner}
\orcid{0009-0000-4360-3731}
%\email{madwagner@tesla.com}%{mwagne22@nd.edu}

\author{Elizabeth Medina}
\orcid{0000-0002-8232-1566}
%\email{emedina0137@gmail.com}%{emedina02@saintmarys.edu}

\affiliation{%
  \institution{University of Notre Dame}
  \city{South Bend}
  %\state{Indiana}
  \country{United States}
  }
  
\author{Juan Garduño Espinosa} 
\orcid{0000-0002-3000-4948}
%\email{jgardunoe@gmail.com}

\author{Horacio Márquez-González}
\orcid{0000-0001-9041-5813}
%\email{horaciohimfg@gmail.com}

\author{Victor Olivar-López}
\orcid{0009-0008-3314-0372}
%\email{vol63@hotmail.com}

\author{Luis E. Juárez-Villegas}
\orcid{0000-0002-9246-7884}
%\email{luisjuarezvillegas@gmail.com}

\author{Martha Avilés-Robles}
\orcid{0000-0001-5995-1503}
%\email{marmtom@hotmail.com}

\author{Elisa Dorantes-Acosta}
\orcid{0000-0001-8992-9315}
%\email{elisadorantes@hotmail.com}

\author{Viridia Avila}
\orcid{0009-0008-6592-2520}
%\email{viridiana.avila25@gmail.com}

\author{Gina Chapa-Koloffon}
\orcid{0000-0001-8726-628X}
%\email{ginachapak@gmail.com}

\author{Elizabeth Cruz}
\orcid{}
%\email{eli.investigacion.him@gmail.com}

\author{Leticia Luis}
\orcid{0009-0008-3909-4338}
%\email{luisleticia83@gmail.com}

\author{Clara Quezada}
\orcid{0009-0003-5277-6994}
%\email{clarq.1973@yahoo.com}

\affiliation{%
  \institution{Hospital Infantil de México Federico Gómez (HIMFG)}
  \city{Mexico City}
  %\state{Mexico}
  \country{Mexico}}
  
\author{Emanuel Orozco}
\orcid{0000-0002-6550-7385}
%\email{emanuel.orozco@insp.mx}

\author{Edson Serván-Mori}
\orcid{0000-0001-9820-8325}
%\email{eservan@insp.mx}
\affiliation{%
  \institution{Instituto Nacional de Salud Pública}
  \city{Mexico City}
  %\state{Mexico}
  \country{Mexico}}

\author{Martha Cordero} 
\orcid{0000-0002-9644-6482}
%\email{marthiux1403@hotmail.com}%{martha.cordero@imp.edu.mx}
\affiliation{%
  \institution{Instituto Nacional De Psiquiatría Ramón De La Fuente Muñiz}
  \city{Mexico City}
  %\state{Mexico}
  \country{Mexico}}

\author{Rubén Martín Payo} 
\orcid{0000-0001-7835-4616}
%\email{martinruben@uniovi.es}
\affiliation{%
  \institution{Universidad de Oviedo, España}
  \city{Oviedo}
  \state{}
  \country{Spain}}

\author{Nitesh V. Chawla}
\orcid{0000-0003-3932-5956}
%\email{nchawla@nd.edu}
\authornote{Corresponding author. Email: nchawla@nd.edu. }
\affiliation{%
  \institution{University of Notre Dame}
  \city{South Bend}
  %\state{Indiana}
  \country{United States}}

\renewcommand{\shortauthors}{Schnur et al.}%, Garcia-Martinez, Soga, Badillo-Urquiola, et al.}

%%
%% The abstract is a short summary of the work to be presented in the article.
\begin{abstract}
We developed SaludConectaMX as a comprehensive system to track and understand the determinants of complications throughout chemotherapy treatment for children with cancer in Mexico. SaludConectaMX is unique in that it integrates patient clinical indicators with social determinants and caregiver mental health, forming a social-clinical perspective of the patient's evolving health trajectory. The system is composed of a web application (for hospital staff) and a mobile application (for family caregivers), providing the opportunity for cooperative patient monitoring in both hospital and home settings. This paper presents the system's preliminary design and usability evaluation results from a 1.5-year pilot study. Our findings indicate that while the hospital web app demonstrates high completion rates and user satisfaction, the family mobile app requires additional improvements for optimal accessibility; statistical and qualitative data analysis illuminate pathways for system improvement. Based on this evidence, we formalize suggestions for health system development in LMICs, which HCI researchers may leverage in future work. 
\end{abstract}


\begin{CCSXML}
<ccs2012>
<concept>
<concept_id>10010405.10010444.10010447</concept_id>
<concept_desc>Applied computing~Health care information systems</concept_desc>
<concept_significance>500</concept_significance>
</concept>
<concept>
<concept_id>10003120.10003121.10003129</concept_id>
<concept_desc>Human-centered computing~Interactive systems and tools</concept_desc>
<concept_significance>500</concept_significance>
</concept>
</ccs2012>
\end{CCSXML}

\ccsdesc[500]{Applied computing~Health care information systems}
\ccsdesc[500]{Human-centered computing~Interactive systems and tools}

\keywords{Health Equity, Mobile Health (mHealth), Pediatric Cancer, Childhood Cancer, Mobile App, Web App}

%\received{23 May 2024}
%\received[revised]{30 July 2024}

\maketitle

\section{Introduction}

Cancer is a rising public health issue across the globe, impacting countries across the Human Development Index (HDI) spectrum \cite{me2024global}. However, more than two-thirds of the world’s cancers will occur in low- and middle-income countries (LMICs) by 2024 \cite{GlobalBurden}. On top of this, the fragmentation of health services in LMICs generates delays in diagnosis and treatment for children with cancer, reducing chances of survival \cite{siqueira2021impacts}. %In Mexico, a middle-income country with cancer as the 3rd leading cause of death \cite{mohar2017cancer}, the national incidence in youth between $0-19$ years old is estimated to be 16 per 100,000, with a corresponding mortality rate of 5 per 100,000; as a result, Mexico is ranked 5th and 6th for pediatric cancer incidence and mortality, respectively, out of 32 Latin American countries \cite{WHO_cancer}. 
Mexico, a middle-income country with cancer as the 3rd leading cause of death \cite{mohar2017cancer}, is ranked 5th and 6th for pediatric cancer incidence (16 per 100,000) and mortality (5 per 100,000), respectively, out of 32 Latin American countries \cite{WHO_cancer}. Meanwhile, Mexico faces significant challenges in the health system. For example, over $56\%$ of youth do not have access to any type of social security program \cite{stat2019}. Moreover, significant diagnosis and treatment delays have been well documented in vulnerable groups of the population %, especially in rural and suburban areas 
\cite{knaul2023setbacks}. %Hospital A offers medical care to children from families without social protection, providing treatment and medical care mainly to children with cancer from the central and southern areas of the Mexican territory, making this institution a relevant context for this work.
In this context, mobile health (mHealth) tools \cite{mccool2022mobile} may play a pivotal role in providing better healthcare access and cooperative health monitoring opportunities within patients' care teams; specifically, facilitating real-time data flow between caregivers and healthcare providers may inform prompt decisions that minimize the risk of complications%during treatment cycles
, which can save lives and reduce financial burden for both health systems and families \cite{lall2018models}.

Often, mHealth tools serve as \textit{supplementary} resources to complement in-person clinic visits and address diverse patient needs \cite{gajarawala2021telehealth}, especially for those in under-served areas \cite{romain2022effect} or with socioeconomic limitations \cite{barwise2023perceptions}. %; by using virtual platforms, patients in low-income and remote regions can receive prompt consultations, mitigating outcome discrepancies across socioeconomic backgrounds \cite{barwise2023perceptions}. 
These tools help bridge health gaps by improving healthcare access, alleviating financial and transportation burdens, enhancing health education, addressing cultural and linguistic sensitivities, and offering culturally relevant interfaces \cite{riley2012health}. They also promote the adoption of digital health records and standardized protocols, %facilitating smooth information exchange among various healthcare facilities, 
improving continuity of care and associated clinical decision-making \cite{haleem2021telemedicine}. While evidence shows that mHealth systems were successfully implemented during the COVID-19 pandemic in LMICs like India, China, Brazil, and Pakistan \cite{GlobalBurden, Bhat2021, Capeleti2021}, none facilitate the care of children with cancer. %, especially those at risk of developing complications after treatment. 
Meanwhile, within the Human-Computer Interaction (HCI) community, several research studies have been conducted in related areas like breast cancer \cite{Jacobs2016, Jacobs2018}, and chronic illnesses \cite{Jacobs2019} more generally; however, these are focused on adult patients living outside of LMIC contexts. Previous HCI research on designing technology for childhood cancer, specifically, has focused on promoting positive experiences during treatment \cite{Choi2020} and identifying and supporting patient needs within the parent-child dyad \cite{Seo2021}. While pediatric cancer care coordination using communication technology has been studied \cite{Nikkhah2022}, this work focuses on balancing role responsibilities and tasks within families, rather than supporting patient monitoring between caregivers and healthcare providers. To the best of our knowledge, no comprehensive tool has been implemented toward this end.

In this work, we partnered with Hospital Infantil de México Federico Gómez (HIMFG) and Instituto Nacional de Salud Pública (INSP) to develop the SaludConectaMX system as a holistic approach to healthcare access and cooperative monitoring for children with cancer in Mexico. %This collaboration is focused on dynamically monitoring complications associated with chemotherapy, which represent the leading cause of death during treatment. Severe neutropenia (absolute neutrophils below 500 cells/mm3) with fever is the most severe complication because it increases the risk of septic shock, and the time frame to start treatment is short (less than 1 hour) \cite{punnapuzha2023febrile}. 
%The goal of this mHealth tool is to support
This mHealth tool supports 
clinicians and families collaboratively track signs and symptoms of complications, %(e.g., febrile neutropenia, persistent neutropenia, neutropenic enterocolitis, nosocomial infections, bleeding, and death), 
social determinants of health (SDOH), and caregiver mental health throughout the patient's treatment. %These latter components are not typically included in clinical prediction models but may contribute to complication risk \cite{}, especially as they evolve over the patient trajectory. 
This paper presents a preliminary design of our system (\textbf{artifact contribution}) and key findings from a 1.5-year pilot study (\textbf{empirical contribution}). The evidence demonstrates that SaludConectaMX has the potential to be a trusted and useful system in the Mexican context, as long as certain design and accessibility factors are addressed. The insights provided in this paper may be leveraged by HCI researchers for mHealth development, particularly within LMICs. % (\textbf{theoretical contribution}).
\section{System Design}

\begin{figure*}
    \centering
    \begin{subfigure}{.45\linewidth}
        \centering
        \includegraphics[width=.98\linewidth]{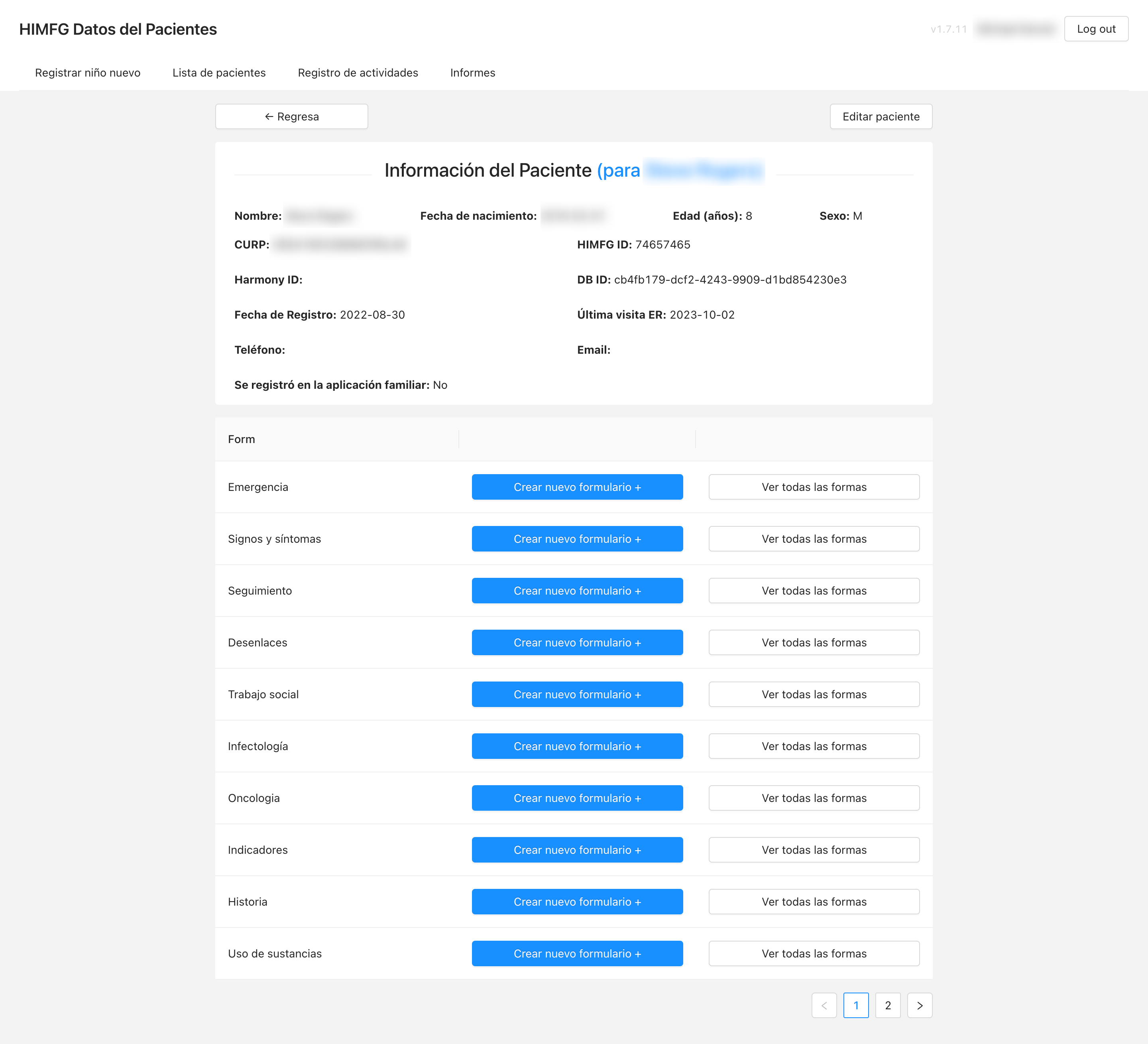}%{Figures/HigherResolution/PatientDashHighRes.png}
        \caption{Hospital web app (HWA) \\ patient form dashboard.}
        \label{fig:HWA_dash}
    \end{subfigure}%    
    \centering
    \begin{subfigure}{.45\linewidth}
        \centering
        \includegraphics[width=0.85\linewidth]{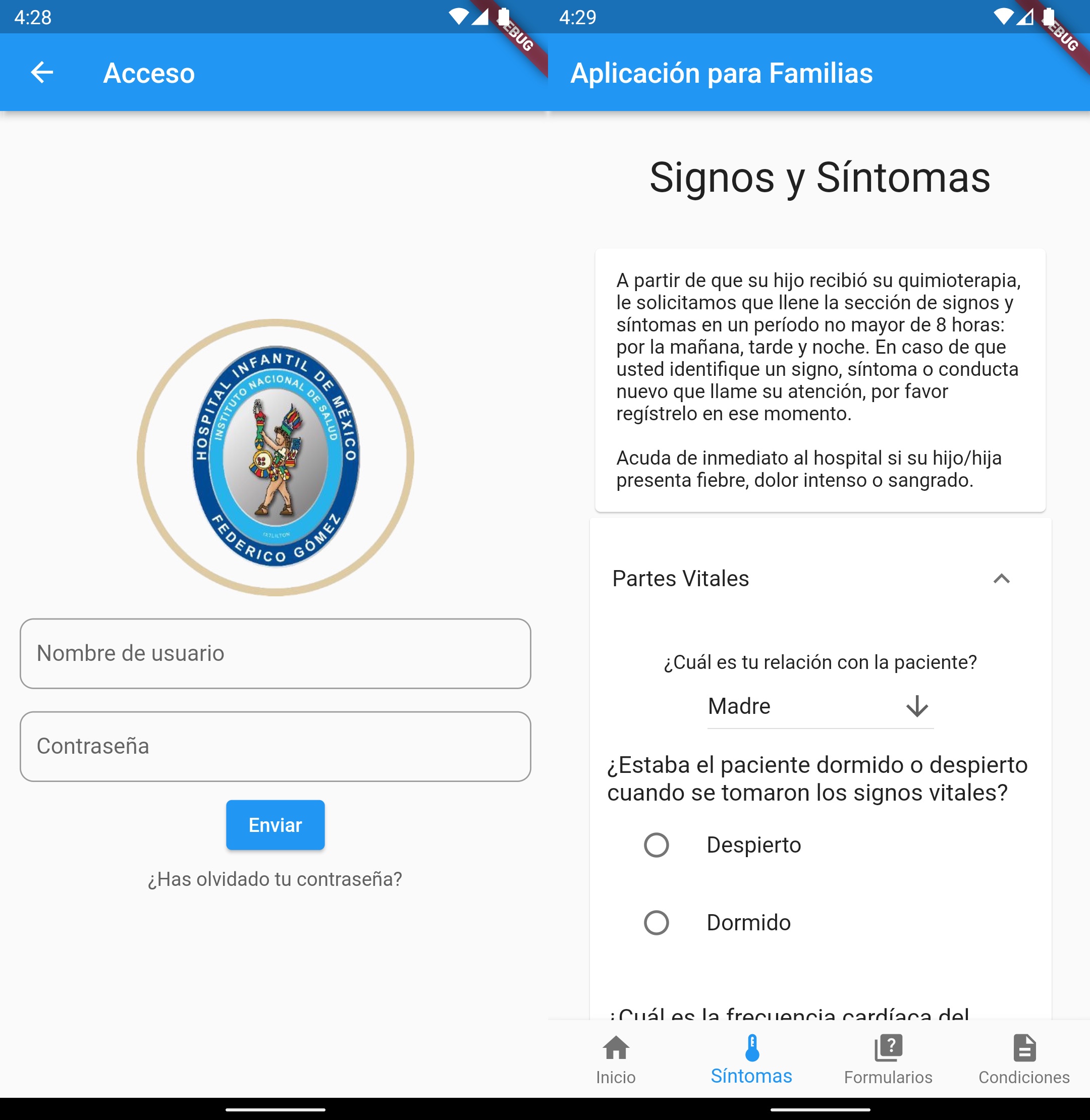}%{Figures/HigherResolution/FMA_HighRes.jpg}
        \caption{Family mobile app (FMA) login page \\ and ``Signs \& Symptoms'' form.}
        \label{fig:FMA_login}
    \end{subfigure}
    
    \caption{Visuals of the SaludConectaMX Hospital Web Application (left) and Family Mobile Application (right).}
    \Description{An image displaying the SaludConectaMX Hospital Web application. On the left is an image of a patient's main dashboard within the HWA and on the right are images of the FMA login page and Signs \& Symptoms form.}
    \label{fig:interfaces}
\end{figure*}

\begin{figure*}
  \centering
  \includegraphics[width=\linewidth]{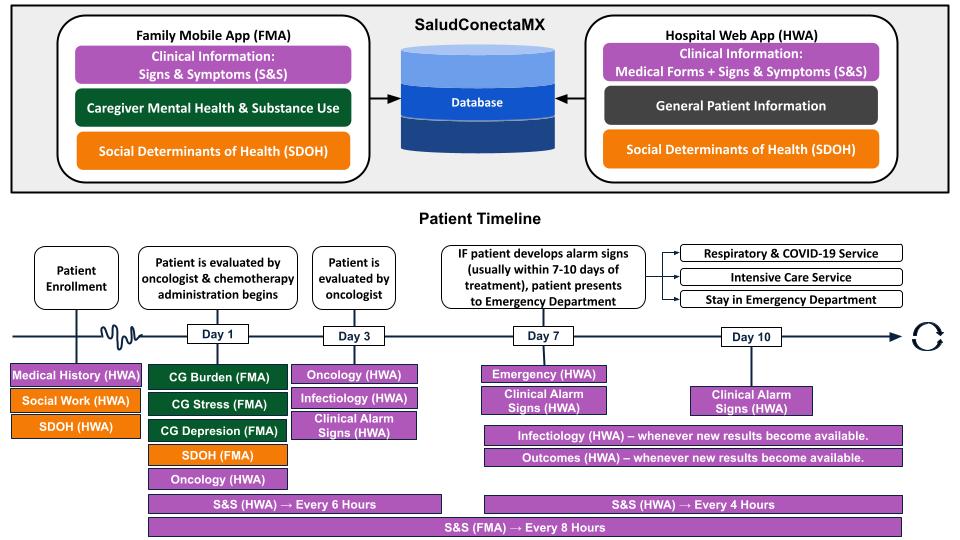}%{Figures/SaludConectaMX_System.jpeg}%{Figures/SuladConectaMX}
  \caption{SuladConectaMX system organization (top) and documentation throughout the patient trajectory (bottom). Colors indicate matchings between high-level data categories captured by the system and relevant forms completed over time. %Note: the FMA provides real-time feedback to caregivers when alarm signs and symptoms arise.
  }%{Interactivity (information exchange) between the Hospital Web Application (left) and the Family Mobile Application (Right)}
  \Description{A figure displaying the organization of the system (top) and how data is collected throughout the patient trajectory (bottom).}
  \label{fig:system}
\end{figure*}

The SuladConectaMX system tracks clinical indicators (i.e., standard medical information, treatment, signs and symptoms of complications, etc.), caregiver attributes (i.e. mental health and substance use), and social determinants of health (SDOH) of pediatric oncology patients served by HIMFG over time. The system consists of two main components: (1) the hospital web application (HWA) for data collection by clinical and social services staff in the hospital setting and (2) the family mobile application (FMA) for data collection by patient caregivers in the home setting and during hospitalizations. Visuals of both interfaces are shown in Figure \ref{fig:interfaces}. Meanwhile, Figure \ref{fig:system} displays the interactions between all components of the system. To follow, we describe these system components and how data is collected and shared between them.
%0.98, 0.85

\subsection{Hospital Web Application} %(For Clinical Staff)}

The HWA is designed for hospital staff (e.g., physicians, social workers, psychiatrists, etc.) to collect pertinent information about the patient in the clinical setting (i.e., during an outpatient oncology appointment, an emergency department visit, and throughout hospitalization). Hospital staff members sign into the HWA via username and password, bringing them to the landing dashboard where all logged patients are listed (Figure \ref{fig:interfaces}). Staff are able to access a particular patient record via the search bar at the top of the page. If the patient is not listed in the dashboard, the staff can navigate to the \textit{Registrar Niño Nuevo} (Register A New Child) tab to add the patient to the system. Each patient’s record consists of a series of electronic forms, which were adapted from the existing paper data collection tools used by HIMFG staff; clicking the \textit{Crear nuevo formulario} (Create a new form) button allows hospital staff to add a new form entry for a patient. The following information is collected through the HWA:

\begin{itemize}
    \item \textbf{General Patient Information}: Family demographics, beneficiary information, and housing characteristics.
    \item \textbf{Clinical Information}: Patient medical history, oncology treatments, emergency visits, alarm signs and symptoms, infection information, and clinical outcomes.
    \item \textbf{SDOH}: Social \& economic conditions, e.g., food security, expenses, difficulties receiving medical attention, etc.
    %\item \textbf{Outcomes}: Upon patient discharge from hospitalization, a hospital staff member records the clinical outcomes that the patient experienced during his or her stay. Outcomes of interest include sepsis, septic shock, bleeding, neutropenic enterocolitis, persistent neutropenia, and death.
    
\end{itemize}

\subsection{Family Mobile Application} %(For Patient Caregivers)}

The FMA is designed with three main goals in mind: (1) to monitor the patient’s clinical condition both at home and throughout hospitalization for potential complications post-chemotherapy treatment, (2) to assess the mental health and substance use patterns of the primary caregivers, and (3) contemporaneous sharing of SDOH variables by the primary caregiver. Families download the app at the hospital with the support of the social work team, who provide a short training using infographic materials. Support service is also provided by engineers from Monday to Saturday: 8:00 A.M. to 8:00 P.M. The FMA includes the following features:

%\begin{figure}[h]
%  \centering
%  \includegraphics[width=.2\columnwidth]{Login.jpg}
%  \includegraphics[width=.201\columnwidth]{Signos.jpg}
%  \caption{Screenshots of the Family Mobile App: Log-in Page (left) and Signs %\& Symptoms Questionnaire (Right)}
%  \Description{A figure displaying }
%\end{figure}

\begin{itemize}
    \item \textbf{Inicio (Home)}: At the home page, caregivers can contact hospital app support by clicking the \textit{WhatsApp} button. This sends a WhatsApp message to the IT team who assists with any technical issue.%, and how to change their password or log out of the app.

    \item \textbf{Síntomas (Symptoms)}: Caregivers monitor the patient at home using the signs and symptoms form. Signs include but are not limited to heart rate, temperature, respiratory rate, attitude/behavior, etc. The app is programmed to alert the user to go to the Emergency Room if certain symptoms arise.%any of the following conditions are met: drowsiness, fever greater than or equal to $38$ degrees Celsius, abdominal pain, diarrhea, bleeding, difficulty breathing, or the heart and respiratory rates are outside the acceptable range for their age.

    \item \textbf{Formularios (Questionnaires)}: Under this tab, caregivers will find two questionnaires. The \textit{SDOH} questionnaire collects data about the family’s social and economic conditions that affect their ability to receive care. The \textit{bienestar del cuidador} (mental health of the primary caregiver) questionnaire collects data about stress, depression, substance use, and burden via the Perceived Stress Scale (PSS) \cite{ramirez2007factor}, Center for Epidemiological Studies-Depression Scale (CES-D-7) \cite{herrero2007brief}, World Health Organization's ASSIST tool \cite{ASSIST-WHO}, and Zarit Burden Interview (ZBI) \cite{bedard2001zarit}, respectively.
    
    \item \textbf{Condiciones (Study Information)}: Since the information collected from families is used for research purposes, caregivers can refer to the study information under this tab. %Caregivers’ mental health information is shared with the hospital staff, who can assist in confirming the diagnosis, provide infographic materials and workshops and referrals to specialists for severe cases of depression.
    
\end{itemize}

%\subsection{Technical Implementation}
%Our web and mobile applications share a common API backend written in Python as a Flask web service hosted in a secure AWS EC2 instance configured to host patient-identifiable information. Access to the EC2 instance is restricted using Okta authentication via Amazon AppStream. Calls to the API are further secured by ensuring the call originated from a logged in user from the web or mobile app. We use PostgreSQL for our database which stores all form data across the patient trajectory (Figure \ref{fig:system}). The patient-facing mobile portion of the application was built with the cross-platform framework Flutter using the Dart programming language and communicates with the same Flask API as the hospital web application. We chose Flutter to be able to write applications for both iOS and Android. 

\subsection{Technical Implementation}
Our web and mobile applications share a common API backend written in Python as a Flask web service hosted in a secure AWS EC2 instance configured to host patient-identifiable information. Access to the EC2 instance is restricted using Okta authentication via Amazon AppStream. Calls to the API are further secured by ensuring the call originated from a logged in user from the web or mobile app. We use PostgreSQL for our database which stores all form data across the patient trajectory (Figure \ref{fig:system}). The patient-facing mobile portion of the application was built with the cross-platform framework Flutter using the Dart programming language and communicates with the same Flask API as the HWA. We chose Flutter to be able to write applications for both iOS and Android.
\section{Pilot Study Findings}

We conducted a preliminary usability evaluation of the SaludConectaMX system focusing on two key elements: (1) the success of practical system use and (2) user satisfaction via surveys and interviews. %We discuss the data collected and findings next. 

\subsection{System Use}

Between June 6th, 2022 and November 21st, 2023 (17.53 months), 310 patients presented to HIMFG for oncology services. Of these patients, 274 (88.4\%) enrolled in the SaludConectaMX system, 21 (6.8\%) rejected participation, 12 (3.9\%) lacked internet access, 6 (1.9\%) were lost during follow-up, and 3 (0.97\%) lacked appropriate technology. Of the 274 enrolled patients, 165 (60.2\%) were male and 109 (39.8\%) were female. The age range of patients at enrollment was between 0.10 and 19.76 years (IQR: 4.95 - 12.34 years). 

System usage results (i.e., the percentage of patients with at least one documented required form type at the relevant time points) are shown in Table \ref{tab:usage_tab}. %, and patient enrollment, separated out into key milestones, can be found in Supplementary Figure \ref{fig:enroll}. 
Over the course of the pilot phase, HWA forms achieved completion rates between $88.2\% - 100\%$, indicating successful use of the system by healthcare professionals. Meanwhile, caregiver form completion rates within the FMA are notably lower ($57.7\% - 63.9\%$), with $162$ ($59.1\%$) patients missing at least one out of six FMA forms. %This challenge was noted soon after the FMA was deployed. 
In order to account for potential barriers, a multidisciplinary team (consisting of medical residents, social workers, and psychologists) began to assist caregivers in downloading the app and completing forms (starting March, 2023); however, there was no statistical difference in the average number of FMA forms completed per patient pre- vs. post-assistance ($p = 0.815$, via permutation test), motivating further investigation into observed missingness. Initial hypotheses included:

\begin{enumerate}
    %\item \textbf{Deploying the FMA 4 months after the HWA.} Upon statistical investigation (permutation tests), we do not observe a significant difference in the average number of FMA forms completed between patients enrolled pre- vs. post- FMA deployment ($p=0.843$); this finding remains consistent after excluding patients who enrolled after healthcare staff began to assist with FMA completion ($p=0.797$). Similarly, we do not find that help from healthcare staff significantly contributed to lower average missingess per patient ($p=0.815$), even after excluding patients who enrolled before FMA deployment ($p=0.546$). This finding motivates the need to investigate additional reasons for missingness.
    
    \item \textbf{Caregiver sensitivities toward reporting poor mental health.} For each of the mental health forms (depression, stress, and burden), we compared the mean scores observed in our population of caregivers to the scores observed in the populations in which the original surveys were developed. We do not find a statistically significant difference in means for stress ($p=0.567$) or burden ($p=0.199$); however, a significant difference for depression scores was noted ($p < 0.001$), as well as strong right-skew ($p=0.008$), which may indicate under-reported depression cases (i.e., data may be ``missing not at random''). Additional support from Psychiatric specialists may mitigate this missingness in the future.
    
    \item \textbf{Social, economic, \& demographic characteristics of the caregivers.} Using observed variables  within the HWA (SDOH, History, and Social Work forms), we investigate associations with FMA missingness. In addition to various associations with caregiver occupations, marital status, and roles, there were statistically significant differences in the mean number of FMA forms completed for those living at residences with and without internet access $(p = 0.019)$, gas $(p=0.024)$, and street lighting $(p=0.048)$. Similarly, the mother's education level was correlated with the number of FMA forms completed $(p=0.012)$. These findings illuminate the need to design a more accessible system. %for caregivers of diverse backgrounds. %Other factors not collected within the HWA that may contribute to missingness (furthered detailed in Section \ref{section:satisfaction}) include limited time, health literacy, and app experience.
    
\end{enumerate}

%Other factors that may contribute to missingness, as reported by caregivers (but not collected in the HWA), include limited time, health literacy, and app experience. %Therefore, there are remaining trade-offs to consider regarding accessibility of the FMA, along with the balance between documentation burden and the quality of future risk prediction models and the benefits they may bring the care team.

\begin{table}
\centering
\caption{System usage results, summarized for each form and system section. Cells are highlighted using a continuous red-yellow-green gradient associated with a $0-100\%$ completion scale. Note: the denominator $N$ drops from $274$ to $169$ for forms that are only required following Emergency visits.}
\label{tab:usage_tab}
\resizebox{\columnwidth}{!}{%
\begin{tabular}{|c|c|c|c|}
%\hline
\rowcolor[HTML]{EFEFEF} 
\hline
\textbf{\begin{tabular}[c]{@{}c@{}}Form Name\\\end{tabular}}          & \textbf{\begin{tabular}[c]{@{}c@{}}Completion Time\\\end{tabular}}                                                         & \textbf{N} & \textbf{\begin{tabular}[c]{@{}c@{}}Completed at \\ least once (\%)\end{tabular}} \\ \hline
\rowcolor[HTML]{FFFFFF} 
Medical History (HWA)                                                 & Enrollment (1x)                                                                                                             & 274        & \cellcolor[HTML]{57BB8A}274 (100\%)                                              \\ \hline
\rowcolor[HTML]{FFFFFF} 
Social Work (HWA)                                                     & \begin{tabular}[c]{@{}c@{}}Enrollment \\ (plus annually)\end{tabular}                                                       & 274        & \cellcolor[HTML]{A1C77A}250 (95.1\%)                                             \\ \hline
\rowcolor[HTML]{FFFFFF} 
SDOH (HWA)                                                            & Enrollment (1x)                                                                                                             & 274        & \cellcolor[HTML]{90C57D}253 (96.2\%)                                             \\ \hline
\rowcolor[HTML]{FFFFFF} 
Oncology (HWA)                                                        & Oncology Appts.                                                                                                             & 274        & \cellcolor[HTML]{93C57D}263 (96.0\%)                                             \\ \hline
\rowcolor[HTML]{FFFFFF} 
\begin{tabular}[c]{@{}c@{}}Caregiver \\ Burden (FMA)\end{tabular}                                                         & Enrollment (1x)                                                                                                             & 274        & \cellcolor[HTML]{F8BC6A}171 (62.4\%)                                             \\ \hline
\rowcolor[HTML]{FFFFFF} 
\begin{tabular}[c]{@{}c@{}}Caregiver \\ Stress (FMA)\end{tabular}                                                         & Enrollment (1x)                                                                                                             & 274        & \cellcolor[HTML]{F8BE6A}175 (63.9\%)                                             \\ \hline
\rowcolor[HTML]{FFFFFF} 
\begin{tabular}[c]{@{}c@{}}Caregiver \\ Depression (FMA)\end{tabular}                                                     & Enrollment (1x)                                                                                                             & 274        & \cellcolor[HTML]{F8BE6A}175 (63.9\%)                                             \\ \hline
\rowcolor[HTML]{FFFFFF} 
\begin{tabular}[c]{@{}c@{}}Caregiver \\ Substance Use (FMA)\end{tabular}                                                   & Enrollment (1x)                                                                                                             & 274        & \cellcolor[HTML]{F8BE6A}175 (63.9\%)                                             \\ \hline
\rowcolor[HTML]{FFFFFF} 
SDOH (FMA)                                                            & Oncology appts.                                                                                                             & 274        & \cellcolor[HTML]{F6B86B}158 (57.7\%)                                             \\ \hline
\rowcolor[HTML]{FFFFFF} 
S\&S (FMA)                                                            & Every 8 Hours                                                                                                               & 274        & \cellcolor[HTML]{F7BB6A}165 (60.2\%)                                             \\ \hline
\rowcolor[HTML]{FFFFFF} 
Emergency (HWA)                                                       & \begin{tabular}[c]{@{}c@{}}Emergency Visits\\ (hospitalization day 1)\end{tabular}                                          & 274        & \cellcolor[HTML]{C0C0C0}169 (61.7\%)                                             \\ \hline
\rowcolor[HTML]{FFFFFF} 
Infectiology (HWA)                                                    & \begin{tabular}[c]{@{}c@{}}At least once during\\ hospitalizations \\ (+ optional during\\  oncology appts.)\end{tabular}   & 169        & \cellcolor[HTML]{A7C879}160 (94.7\%)                                             \\ \hline
\rowcolor[HTML]{FFFFFF} 
\begin{tabular}[c]{@{}c@{}}Clinical Alarm \\ Signs (HWA)\end{tabular} & \begin{tabular}[c]{@{}c@{}}Days 1 and 3 of \\ hospitalizations\\ (+ optional during\\  oncology appts.)\end{tabular}        & 169        & \cellcolor[HTML]{F7D567}151 (89.3\%)                                             \\ \hline
\rowcolor[HTML]{FFFFFF} 
S\&S (HWA)                                                            & \begin{tabular}[c]{@{}c@{}}Every 6 hours during\\ Oncology appts. \&\\ Every 4 hours during\\ hospitalizations\end{tabular} & 169        & \cellcolor[HTML]{B9CB75}158 (93.5\%)                                             \\ \hline
\rowcolor[HTML]{FFFFFF} 
Outcomes (HWA)                                                        & \begin{tabular}[c]{@{}c@{}}Once during \\ hospitalizations\end{tabular}                                                     & 169        & \cellcolor[HTML]{FED567}149 (88.2\%)                                             \\ \hline
\end{tabular}%
}
\end{table}

\subsection{User Satisfaction}
\label{section:satisfaction}

User satisfaction was studied in two ways. First, healthcare personnel were asked to complete the Mobile Application Rating Scale (MARS) survey \cite{terhorst2020validation, payo2019spanish}, which measures an health app’s engagement, functionality, aesthetics, and information quality, as evaluated by the user on 5-point rating scales. This tool has been previously validated for pediatric oncology patients in the Spanish context \cite{amor2020assessing, narrillos2022mobile}, which translated well to the Mexican context without adaptation. %, making it the most appropriate option to evaluate our system, to the best of our knowledge.
Then, our research team selected a small group of healthcare personnel and caregivers to participate in unstructured interviews regarding their experiences with the system. The findings are summarized as follows.  

\subsubsection{MARS Survey Results} 
Between July 17, 2023 and October 12, 2023, 18 health personnel completed the MARS survey, of which 4 (22.2\%) were male and 14 (77.8\%) were female. The average age of the survey respondents was 24.4 years (IQR: 24.0 - 25.0). Of the 18 respondents, 1 (5.6\%) completed their master’s degree, 1 (5.6\%) has an incomplete master’s degree, 8 (44.4\%) completed their bachelor’s degree, 6 (33.3\%) have an incomplete bachelor’s degree, and 2 (11.1\%) completed high school. The survey results are shown in Table \ref{tab:MARS_table}. 

For most survey subcategories, the average rating was at least 3.33 (median at least 4.0), exhibiting acceptable-to-good quality system design. In particular, ``Credibility,'' ``Navigation,'' and ``Information Quality,'' all averaged above a 4.0 rating, emphasizing the professional and comprehensive nature of the HWA, which inspires confidence that future data analysis will provide meaningful insights to the care team. The exceptions were the “Engagement” subcategories labeled “Fun,” “Individual Adaptability,” and “Interactivity.” While engagement is important for general system design, we argue that including “Fun” aspects may not be appropriate within a health monitoring system for chronically ill children. Similarly, “Individual Adaptability” was not a priority for this system, as standardized data collection was required across all patients. However, “Interactivity” components, such as notifications, reminders, or sharing options, may improve communication and data quality, and we plan to implement these features in future work. %Meanwhile, the top three MARS subcategories were ``Credibility,'' ``Navigation,'' and ``Information Quality,'' all of which averaged above a 4.0 rating. These results emphasize the professional and comprehensive nature of the HWA and inspire confidence that future data analysis will provide meaningful insights to the care team. %In future work, we plan to evaluate user satisfaction for the family mobile app. 

\begin{table}
\centering
\caption{Hospital Web App for Health Personnel, user satisfaction (MARS) survey results, summarized within survey subcategories across respondents. Cells are colored using a continuous red-yellow-green gradient associated with the 1-5 rating scale values (1 = inadequate; 5 = excellent).}
\label{tab:MARS_table}
\resizebox{\columnwidth}{!}{%
\begin{tabular}{|c|c|c|c|
>{\columncolor[HTML]{ABC978}}c |c
>{\columncolor[HTML]{ABC978}}c |}
\hline
\cellcolor[HTML]{EFEFEF}\textbf{\begin{tabular}[c]{@{}c@{}}MARS \\ Category\end{tabular}} & \cellcolor[HTML]{EFEFEF}\textbf{\begin{tabular}[c]{@{}c@{}}MARS \\ Subcategory\end{tabular}} & \cellcolor[HTML]{EFEFEF}\textbf{N} & \cellcolor[HTML]{EFEFEF}\textbf{Mean} & \cellcolor[HTML]{EFEFEF}\textbf{Median} & \multicolumn{2}{c|}{\cellcolor[HTML]{EFEFEF}\textbf{IQR}}                       \\ \hline
                                                                                          & \textbf{Graphics}                                                                            & 18                                 & \cellcolor[HTML]{BACB74}3.83          & 4                                       & \multicolumn{1}{c|}{\cellcolor[HTML]{EAD36A}3.25} & \cellcolor[HTML]{6CBF85}4.75 \\ \cline{2-7} 
                                                                                          & \textbf{Layout}                                                                              & 18                                 & \cellcolor[HTML]{DBD16D}3.44          & 4                                       & \multicolumn{1}{c|}{\cellcolor[HTML]{F2A96D}2}    & 4                            \\ \cline{2-7} 
\multirow{-3}{*}{\textbf{Aesthetics}}                                                     & \textbf{\begin{tabular}[c]{@{}c@{}}Visual \\ Appeal\end{tabular}}                            & 18                                 & \cellcolor[HTML]{C7CD72}3.67          & 4                                       & \multicolumn{1}{c|}{\cellcolor[HTML]{ABC978}4}    & 4                            \\ \hline
                                                                                          & \textbf{Fun}                                                                                 & 18                                 & \cellcolor[HTML]{F3AD6C}2.11          & \cellcolor[HTML]{F2A96D}2               & \multicolumn{1}{c|}{\cellcolor[HTML]{E67C73}1}    & \cellcolor[HTML]{FFD666}3    \\ \cline{2-7} 
                                                                                          & \textbf{\begin{tabular}[c]{@{}c@{}}Individual \\ Adaptability\end{tabular}}                  & 18                                 & \cellcolor[HTML]{F0A16E}1.83          & \cellcolor[HTML]{E67C73}1               & \multicolumn{1}{c|}{\cellcolor[HTML]{E67C73}1}    & \cellcolor[HTML]{FFD666}3    \\ \cline{2-7} 
                                                                                          & \textbf{Interactivity}                                                                       & 18                                 & \cellcolor[HTML]{F8BC6A}2.44          & \cellcolor[HTML]{FFD666}3               & \multicolumn{1}{c|}{\cellcolor[HTML]{F2A96D}2}    & \cellcolor[HTML]{FFD666}3    \\ \cline{2-7} 
                                                                                          & \textbf{Interest}                                                                            & 18                                 & \cellcolor[HTML]{CCCE70}3.61          & \cellcolor[HTML]{FFD666}3               & \multicolumn{1}{c|}{\cellcolor[HTML]{FFD666}3}    & \cellcolor[HTML]{6CBF85}4.75 \\ \cline{2-7} 
\multirow{-5}{*}{\textbf{Engagement}}                                                     & \textbf{\begin{tabular}[c]{@{}c@{}}Target \\ Group\end{tabular}}                             & 18                                 & \cellcolor[HTML]{B1CA76}3.94          & 4                                       & \multicolumn{1}{c|}{\cellcolor[HTML]{EAD36A}3.25} & 4                            \\ \hline
                                                                                          & \textbf{\begin{tabular}[c]{@{}c@{}}Gestural \\ Design\end{tabular}}                          & 18                                 & \cellcolor[HTML]{E4D26B}3.33          & 4                                       & \multicolumn{1}{c|}{\cellcolor[HTML]{F5B46B}2.25} & 4                            \\ \cline{2-7} 
                                                                                          & \textbf{Navigation}                                                                          & 18                                 & \cellcolor[HTML]{A2C879}4.11          & 4                                       & \multicolumn{1}{c|}{\cellcolor[HTML]{ABC978}4}    & \cellcolor[HTML]{6CBF85}4.75 \\ \cline{2-7} 
                                                                                          & \textbf{Performance}                                                                         & 18                                 & \cellcolor[HTML]{BACB74}3.83          & 4                                       & \multicolumn{1}{c|}{\cellcolor[HTML]{EAD36A}3.25} & 4                            \\ \cline{2-7} 
\multirow{-4}{*}{\textbf{Functionality}}                                                  & \textbf{Usability}                                                                           & 18                                 & \cellcolor[HTML]{D0CF70}3.56          & 4                                       & \multicolumn{1}{c|}{\cellcolor[HTML]{FFD666}3}    & 4                            \\ \hline
                                                                                          & \textbf{\begin{tabular}[c]{@{}c@{}}Description \\ Accuracy\end{tabular}}                     & 18                                 & \cellcolor[HTML]{B5CA76}3.89          & 4                                       & \multicolumn{1}{c|}{\cellcolor[HTML]{ABC978}4}    & 4                            \\ \cline{2-7} 
                                                                                          & \textbf{Credibility}                                                                         & 18                                 & \cellcolor[HTML]{94C57D}4.28          & 4                                       & \multicolumn{1}{c|}{\cellcolor[HTML]{ABC978}4}    & \cellcolor[HTML]{6CBF85}4.75 \\ \cline{2-7} 
                                                                                          & \textbf{\begin{tabular}[c]{@{}c@{}}Evidence \\ Base\end{tabular}}                            & 10                                 & \cellcolor[HTML]{D5D06F}3.5           & 4                                       & \multicolumn{1}{c|}{\cellcolor[HTML]{FFD666}3}    & 4                            \\ \cline{2-7} 
                                                                                          & \textbf{Goals}                                                                               & 16                                 & \cellcolor[HTML]{C0CC73}3.75          & 4                                       & \multicolumn{1}{c|}{\cellcolor[HTML]{FFD666}3}    & 4                            \\ \cline{2-7} 
                                                                                          & \textbf{\begin{tabular}[c]{@{}c@{}}Info \\ Quality\end{tabular}}                             & 18                                 & \cellcolor[HTML]{A6C879}4.06          & 4                                       & \multicolumn{1}{c|}{\cellcolor[HTML]{ABC978}4}    & \cellcolor[HTML]{6CBF85}4.75 \\ \cline{2-7} 
                                                                                          & \textbf{\begin{tabular}[c]{@{}c@{}}Info \\ Quantity\end{tabular}}                            & 17                                 & \cellcolor[HTML]{B6CB75}3.88          & 4                                       & \multicolumn{1}{c|}{\cellcolor[HTML]{ABC978}4}    & 4                            \\ \cline{2-7} 
\multirow{-7}{*}{\textbf{\begin{tabular}[c]{@{}c@{}}Information \\ Quality\end{tabular}}} & \textbf{\begin{tabular}[c]{@{}c@{}}Visual Info \\ Quality\end{tabular}}                      & 16                                 & \cellcolor[HTML]{B1CA76}3.94          & 4                                       & \multicolumn{1}{c|}{\cellcolor[HTML]{ABC978}4}    & 4                            \\ \hline
\end{tabular}%
}
\end{table}

\subsubsection{Unstructured Interview Results}

 Two members of our research team conducted open-ended interviews to identify barriers experienced by system users. Participants included 1 social worker, 6 medical residents, 1 family who participated in the system, and 1 family who elected not to participate. Detailed findings from these interviews are provided in Table \ref{tab:interviews}. Key problems and actionable change recommendations are summarized as follows:

 \begin{itemize}

     \item \textbf{Additional Resources \& Helpful Functionalities:} Hospital staff noted that inconsistent internet availability within the hospital contributes to form submission difficulties, which may be resolved with additional modems; meanwhile, caregivers noted that limited mobile credit combined with a lack of perceived benefit (often due to limited health and tech literacy) prevented frequent use of the FMA. Participants also expressed that the system %, as it is currently designed, provides few incentives for caregivers to engage. Specifically, the system 
     could benefit from functionalities that produce more direct value and incentivize use, such as plan of care notes, reminders, appointment scheduling, chat bots to resolve doubts \cite{Tseng2023}, emergency event communication, and elimination of factors that contribute to documentation fatigue. These improvements will be made in future work. %modest financial compensation for caregivers. However, a long-term solution for caregivers is needed. 
     
     %\item \textbf{Additional Resources:} Hospital staff noted that inconsistent internet availability within the hospital contributes to form submission difficulties; meanwhile, caregivers noted that limited mobile credit combined with a lack of compensation for data collection efforts prevented frequent use of the FMA. The research team will provide hospital modems and more helpful functionalities that incentivize caregiver. %modest financial compensation for caregivers. However, a long-term solution for caregivers is needed. 
     \item \textbf{Intuitive Design \& Training:} Participants indicated that app usage was not always intuitive (e.g., understanding inclusion criteria, to whom questions pertain, when to submit forms, whether or not forms were properly submitted, etc.). Caregivers noted that these difficulties were compounded by insufficient explanation at the hospital. Therefore, the research team will provide more instructive detail within the system and ensure more robust training for healthcare staff, as well as educational supplements for caregivers, targeting both health and tech literacy.
     
     %\item \textbf{Helpful System Functionalities:} Participants expressed that the system %, as it is currently designed, provides few incentives for caregivers to engage. Specifically, the system 
     %could benefit from functionalities that produce more direct value and incentivize use such as plan of care notes, reminders, appointment scheduling, chat bots to resolve doubts, emergency event communication, and elimination of redundancies and non-applicable prompts to prevent documentation fatigue. These improvements will be made in future work.
 \end{itemize}

\begin{table*}[ht]
\caption{Barrier findings and suggested system improvements from unstructured interviews with healthcare staff and caregivers.}
\label{tab:interviews}
\resizebox{\linewidth}{!}{%
\begin{tabular}{@{}c|lcl@{}}
\toprule
\multicolumn{1}{l|}{\textbf{Interviewee}}                                                             & \textbf{Barrier Description}                                                                                                                                                                                                                                    & \textbf{Barrier Classification}                                                          & \textbf{System Recommendation}                                                                                                                                                                                                                                                                             \\ \midrule
\multirow{5}{*}{\textbf{\begin{tabular}[c]{@{}c@{}}Social \\ Worker\end{tabular}}}                    & \begin{tabular}[c]{@{}l@{}}(HWA) A limited number of healthcare staff are available for \\ data collection; those with little system experience were recruited.\end{tabular}                                                          & Training / Education                                                                     & \begin{tabular}[c]{@{}l@{}}More thorough training for system users will help \\ reduce data collection challenges and biases.\end{tabular}                                                                                                                                                                 \\ \cmidrule(l){2-4} 
                                                                                                      & \begin{tabular}[c]{@{}l@{}}(FMA) Caregivers experience fatigue and confusion when \\ filling out forms repeatedly throughout the system.\end{tabular}                                                                                                           & Design / Functionality                                                                   & \begin{tabular}[c]{@{}l@{}}More user-friendly and instructive design adjustments \\ will be made in future work.\end{tabular}                                                                                                                                                                              \\ \cmidrule(l){2-4} 
                                                                                                      & \begin{tabular}[c]{@{}l@{}}(FMA) Caregivers are less interested in participating if they \\ have had poor experiences / interactions with physicians.\end{tabular}                                                                                              & Training / Education                                                                     & \begin{tabular}[c]{@{}l@{}}Healthcare staff will help build positive rapport with \\families and introduce them to the benefits of the\\ system for their child's wellbeing.\end{tabular}                                                                                                                \\ \cmidrule(l){2-4} 
                                                                                                      & \begin{tabular}[c]{@{}l@{}}(FMA) Caregivers are reluctant to use the app, as they do \\ not perceive any direct benefits.\end{tabular}                                                                                 & \begin{tabular}[c]{@{}c@{}}Design / Functionality \\ + Training / Education\end{tabular} & \begin{tabular}[c]{@{}l@{}}Develop more helpful app features (e.g., reminders, \\appointment scheduling, chatbot), as well as health \\literacy education.\end{tabular}                                                                                                                  \\ \cmidrule(l){2-4} 
                                                                                                      & \begin{tabular}[c]{@{}l@{}}(FMA) Caregivers often answer questions that pertain to \\ the family unit only with respect to the patient (e.g., food \\ security responses often do not reflect sacrifices that \\ parents make for their children).\end{tabular} & \begin{tabular}[c]{@{}c@{}}Design / Functionality\\ + Training / Education\end{tabular}  & \begin{tabular}[c]{@{}l@{}}Adding further specifications for certain form \\ questions will ensure better data quality. These \\ details can also be reinforced through training.\end{tabular}                                                                                                             \\ \midrule
\multirow{4}{*}{\textbf{\begin{tabular}[c]{@{}c@{}}Medical \\ Residents\end{tabular}}}                & \begin{tabular}[c]{@{}l@{}}(HWA + FMA) Internet access is often limited both inside \\ and outside the hospital, making it difficult to submit forms.\end{tabular}                                                                                              & Resources                                                                                & \begin{tabular}[c]{@{}l@{}}Adding modems within the hospital may resolve \\ this issue for the hospital staff.\end{tabular}                                                                                                                                                                                \\ \cmidrule(l){2-4} 
                                                                                                      & \begin{tabular}[c]{@{}l@{}}(FMA) Caregivers often forget to fill out forms and attend \\ appointments. Others only use the app when they perceive \\ that their child is at risk for a complication event.\end{tabular}                                         & \begin{tabular}[c]{@{}c@{}}Design / Functionality \\ + Training / Education\end{tabular} & \begin{tabular}[c]{@{}l@{}}Additional features may assist and incentivize use (e.g., \\reminders, appointment scheduling, chat bots to resolve \\doubts, educational videos to enhance health literacy, etc.).\end{tabular} \\ \cmidrule(l){2-4} 
                                                                                                      & \begin{tabular}[c]{@{}l@{}}(HWA + FMA) Patients and their caregivers often require \\ medical assistance during the journey to the hospital once \\ symptoms of a complication begin.\end{tabular}                                                              & Design / Functionality                                                                   & \begin{tabular}[c]{@{}l@{}}Additional features that support communication \\ between caregivers \& hospital staff will assist with \\ emergency events (e.g., an emergency button to notify \\ the hospital about incoming patients, WhatsApp, etc.).\end{tabular}                                         \\ \cmidrule(l){2-4} 
                                                                                                      & \begin{tabular}[c]{@{}l@{}}(HWA) Some fields are left blank in the emergency forms \\ and are only filled at a later time, introducing data bias.\end{tabular}                                                                                                  & \begin{tabular}[c]{@{}c@{}}Design / Functionality \\ + Training / Education\end{tabular} & \begin{tabular}[c]{@{}l@{}}Instructional design improvements and additional \\ training for healthcare staff may ensure that fields \\ are completed at the proper time points.\end{tabular}                                                                                                               \\ \midrule
\multirow{5}{*}{\textbf{\begin{tabular}[c]{@{}c@{}}Participating \\ Caregivers\end{tabular}}}         & \begin{tabular}[c]{@{}l@{}}(FMA) Caregivers expressed they they feel uncomfortable \\ taking vital signs and do not collect this data as a result.\end{tabular}                                                                                                 & Training / Education                                                                     & \begin{tabular}[c]{@{}l@{}}Additional education and training is needed to \\ support caregivers with at-home monitoring.\end{tabular}                                                                                                                                                                      \\ \cmidrule(l){2-4} 
                                                                                                      & \begin{tabular}[c]{@{}l@{}}(FMA) Caregivers reflect that they did not consider using \\ the app until their child experienced an emergency event.\end{tabular}                                                                                                  & Training / Education                                                                     & \begin{tabular}[c]{@{}l@{}}Improve communication between healthcare staff \\ and families via additional training.\end{tabular}                                                                                                                                               \\ \cmidrule(l){2-4} 
                                                                                                      & \begin{tabular}[c]{@{}l@{}}(FMA) Caregivers express uncertainty as to whether or \\ not data has been properly logged in the system.\end{tabular}                                                                                                               & Design / Functionality                                                                   & \begin{tabular}[c]{@{}l@{}}More intuitive app design is needed to confirm \\ that data has been logged properly by the user.\end{tabular}                                                                                                                                                                  \\ \cmidrule(l){2-4} 
                                                                                                      & \begin{tabular}[c]{@{}l@{}}(FMA) Insufficient mobile credit prevents frequent data \\ collection.\end{tabular}                                                                                                                                                  & Resources                                                                                & \begin{tabular}[c]{@{}l@{}}Focus groups and ethnographic studies may reveal \\ sustainable solutions that ensure accessibility.\end{tabular}                                                                                \\ \cmidrule(l){2-4} 
                                                                                                      & \begin{tabular}[c]{@{}l@{}}(FMA) Most app functionalities are not explained at the \\ time the app is downloaded at the hospital (e.g., some \\ caregivers were unaware of mental health resources \\available within the app).\end{tabular}           & \begin{tabular}[c]{@{}c@{}}Design / Functionality \\ + Training / Education\end{tabular} & \begin{tabular}[c]{@{}l@{}}Caregivers may benefit from an in-app walkthrough \\to see available features. Additional training is also \\ needed to improve communication between healthcare \\staff and caregivers.\end{tabular}                                                                             \\ \midrule
\multirow{3}{*}{\textbf{\begin{tabular}[c]{@{}c@{}}Non- \\ Participating \\ Caregivers\end{tabular}}} & \begin{tabular}[c]{@{}l@{}}(FMA) Caregivers express that they understand what an \\ emergency event looks like and do not believe the app will \\ provide added support.\end{tabular}                                                                           & \begin{tabular}[c]{@{}c@{}}Design / Functionality \\ + Training / Education\end{tabular} & \begin{tabular}[c]{@{}l@{}}Add functionalities that incentivize use of the app \\ beyond emergency event recognition (see above), \\ as well as education to increase health literacy.\end{tabular}                                                                                                                    \\ \cmidrule(l){2-4} 
                                                                                                      & \begin{tabular}[c]{@{}l@{}}(FMA) Caregivers express that instructions for care at \\ home are not always clear. %Written instructions from \\ physicians would be helpful.
                                                                                                      \end{tabular}                                                                          & Design / Functionality                                                                   & \begin{tabular}[c]{@{}l@{}}Provide functionality that allows physicians to record\\ plan of care instructions for caregivers to reference.\end{tabular}                                                                                                                             \\ \cmidrule(l){2-4} 
                                                                                                      & \begin{tabular}[c]{@{}l@{}}(FMA) Some caregivers had no knowledge of the app \\ until late in their child's diagnosis / treatment.\end{tabular}                                                                                                                 & Training / Education                                                                     & \begin{tabular}[c]{@{}l@{}}Provide hospital staff with clarification on patient \\ inclusion criteria for system use.\end{tabular}                                                                                                                                                                         \\ \bottomrule
\end{tabular}%
}
\end{table*}

\section{Conclusion}
SaludConectaMX is a comprehensive pediatric cancer monitoring tool that facilitates collaboration between caregivers and healthcare professionals through dual interfaces. Although successful engineering of mHealth systems in LMICs faces multiple challenges, the work reported in this paper represents good progress toward a practical system that addresses health disparities through technological intervention. %Future system improvements will focus more intently on caregiver expectations and needs.
%Successful engineering of mHealth systems in LMICs faces multiple challenges; %and will require further investigation into the structural, cultural, social, and economic factors that play into observed disparities. 
%and the lessons learned from our pilot phase may inspire future HCI research geared toward similar goals. 
%We distill these lessons into three key takeaways:
We distill the lessons learned from our pilot phase into three key takeaways to inspire future HCI research geared toward similar goals:

\begin{enumerate}
    \item Health monitoring tools must not contribute to care team burden or confusion; instead, consistent use should be incentivized via intuitive design and helpful functionalities.
    \item System roll-out must incorporate extensive training to ensure that healthcare professionals can confidently use the system and communicate instructions to caregivers for successful use at home. Similarly, robust education for caregivers is needed with respect to health and tech literacy.
    
    %\item The lack of accessibility particularly for families with socioeconomic limitations is a major barrier for technological intervention, which can be addressed (in the short-term) by supplying sufficient compensation for participation. However, long-term, sustainable solutions will require additional investigation.
    %\balance
    
    \item Sustainable solutions are needed for long-term adoption of mHealth interventions, especially for those with socioeconomic limitations who stand to benefit the most. Thus, focus groups and ethnographic studies are required to effectively identify barriers and facilitate co-created (re-)design of said systems to optimally serve users from diverse backgrounds. We plan to implement these studies in future work. 
\end{enumerate}

%\balance

\begin{acks}
We thank the Lucy Family Institute for Data and Society, the University of Notre Dame, the National Institute of Pediatrics ‘Hospital Infantil de México Federico Gómez’, and Mexico's National Institute of Public Health for supporting this research. Special thanks to HIMFG staff and the participating patients and their families.
\end{acks}
\bibliographystyle{ACM-Reference-Format}
\balance
\bibliography{00_References}
%\newpage
%\appendix
%\input{8_Supplementary}
\end{document}